\newcommand{\up}{\uparrow} 
\newcommand{\down}{\downarrow} 
\newcommand{\nn}{\nonumber} 
\newcommand{\be}{\begin{align}} 
\newcommand{\ee}{\end{align}} 
\newcommand{\bearr}{\begin{align}\begin{split}} 
\newcommand{\eearr}{\end{split}\end{align}}

\documentclass[twocolumn]{jpsj2} 
\usepackage{psfrag} 
\usepackage{latexsym} 
\usepackage{graphicx}

\title{Dynamical mean field theory of optical third harmonic generation} 
\author{ 
S. Akbar \textsc{Jafari}$^{1,2}$\thanks{E-mail address: akbar@imr.tohoku.ac.jp},  
Takami \textsc{Tohyama}$^{1}$
and Sadamichi \textsc{Maekawa}$^{1,3}$
} 
 
\inst{$^{1}$ Institute for Materials Research, 
Tohoku University, Sendai 980-8577, Japan \\ 
$^{2}$Department of Physics, Isfahan University of Technology, Isfahan 
84156, Iran\\ 
$^{3}$CREST, Japan Science and Technology Agency, Kawaguchi 
332-0012, Japan} 
 
\abst{ 
We formulate the third harmonic generation (THG) within  the
dynamical mean field theory (DMFT) approximation of the Hubbard model.
In the limit of large dimensions, where DMFT becomes exact, 
the vertex corrections to current vertices are identically zero, 
and hence the calculation of the THG spectrum reduces to a time-ordered 
convolution, followd by appropriate analytic continuuation. We present
the typical THG spectrum of the Hubbard model obtained by this method.
Within our DMFT calculation, we observe a nontrivial approximate {\em scaling} function
describing the THG spectra in all Mott insulators, independent of the gap 
magnitude. 
} 
\kword{ 
Third harmonic generation, Mott insulator, Dynamical mean field 
} 
 
\recdate{\today} 
\begin{document} 
\maketitle 
 
   Nonlinear optical interactions of laser fields with matter 
provide powerful spectroscopic tools for the understanding of 
microscopic processes. The ability to control pulse durations 
(to a few femtoseconds), bandwidths (up to 1 Hz resolution), and 
peak intensities (up to $10^{19}$W/cm$^2$) provides novel 
probes of elementary dynamic events of matter\cite{MukamelBook}. 
 
Observation of large third order nonlinear susceptibility in 
a quasi one dimensional 
Mott insulator Sr$_2$CuO$_3$ ($\chi^{(3)}$ values 
in the range $10^{-8}$ to $10^{-5}$ e.s.u.)  
\cite{KishidaNature,KishidaPRL} poses the problem of nonlinear
optical response in correlated insulators.

In systems with large on-site  
Coulomb interaction, the 1D system has the largest optical nonlinearity 
because of the decoupling of spin and charge degrees of 
freedom\cite{Mizuno, HolonDoublon}. On the other hand,  
mean field analysis shows that among 
SDW-ordered systems, the largest third order optical response 
appears in 2D\cite{JafariJPSJ}.  
A natural generalization to higher dimensions is through 
the dynamical mean field theory (DMFT). 
DMFT includes the effect of local quantum fluctuations
and becomes exact in the limit of infinite dimensions.

   In this limit the self-energy becomes a purely local 
quantity determined from a self-consistent Anderson model. 
Then the matrix elements determining spectral weights are  
encoded in the local self-energy. In this limit the matrix element  
summations reduce to density of states (DOS) integrations. Therefore 
the combined DOS and self-energy effects gives rise to nonlinear
optical response of the system.

   We formulate a nonlinear response theory, with example
of THG within DMFT approximation and prescribe a numerically feasible 
method to avoid expensive computations. To our knowledge
this is the first application of DMFT to nonlinear optics.

   The general THG line shape within DMFT framework consists in 
a strong peak at three photon resonance, followed by a shoulder at
two photon resonance, and a very weak feature at one photon resonance.
The three-photon contributions obtained for various on-site Coulomb
repulsion fall approximately on the same curve, if
we scale the frequencies with the gap magnitude.
This behavior is similar to the
one observed in band insulators\cite{JafariJPSJ}, 
where a single particle
picture describes the nonlinear optical processes.
 
   We start with the Hubbard model at half filling 
\begin{small}
\begin{align} 
   H &= \frac{\tilde t}{\sqrt{d}}\! \! \sum_{\langle i,j\rangle,s} \!\!\! 
   \left(c^{\dagger}_{is} c_{js}+{\rm h.c}\right) 
    \! +\!  U \sum_j\! \!  \left(n_{j\up}\!-\!\frac{1}{2}\right) 
    \!\!\left(n_{j\down}\!-\!\frac{1}{2}\right), 
\end{align} 
\end{small}
where $c^{\dagger}_{is}$ creates an electron at site $i$ with 
spin $s=\up,\down$. The dimension of the lattice is $d$. 
Here the $1/\sqrt d$ scaling of the hopping term ensures that 
average kineitc energy per particle in the limit of $d\to\infty$ 
remains finite\cite{KotliarRMP}. 
 
   If we now imagine that we integrate out all degrees of freedom 
on various lattice sites, except for the one at the origin, we will 
be left with an effective action for this "impurity" site: 
\begin{align} 
\begin{split}
    S_{\rm eff} &= -\int_0^\beta\int_0^\beta d\tau d\tau' 
    \sum_\sigma c^\dagger_{o\sigma}(\tau){\cal G}_o^{-1}(\tau-\tau') 
    c_{o\sigma}(\tau') \\ 
    &+U\int_0^\beta n_{o\up}(\tau)n_{o\down}(\tau). 
\end{split}
\end{align} 
Here the impurity propagator ${\cal G}_0(\tau-\tau')$ describes 
temporal quantum fluctuations between the four possible states 
of a single site at the origin, which must be determined 
self-cosistently. 
 
   The simplest way to solve such effective impurity problem is 
the second order perturbation, which gives 
\begin{align} 
   \Sigma(\tau)=\frac{U^2}{4}{\cal G}_o(\tau){\cal G}_o(\tau) 
   {\cal G}_o(-\tau). 
   \label{Sigma.eqn} 
\end{align} 
This gives the lattice Green's function as,
\begin{align} 
    G(\vec k,i\omega_n)&=1/(i\omega_n-\varepsilon_{\vec k}-\Sigma(i\omega_n)).
\end{align} 
The projection of this function on site 'o' is given by
\begin{align} 
   G(i\omega_n) &= \int \frac{D(\varepsilon)d\varepsilon} 
   {i\omega_n-\varepsilon-\Sigma(i\omega_n)} 
   \label{G.eqn} 
\end{align} 
where $D(\varepsilon)$ is the lattice density of states.
Finally the self-consistency between lattice ($G$) and impurity 
(${\cal G}_o$) is via the Dyson equation 
\begin{align} 
   {\cal G}_o^{-1}(i\omega_n)=\Sigma(i\omega_n)+G^{-1}(i\omega_n), 
   \label{Dyson.eqn} 
\end{align} 
which is used to update ${\cal G}_o$ if the consistency has not  
been achieved yet\cite{KotliarRMP}. 
 
   Solving the set of equations (\ref{Sigma.eqn}), (\ref{G.eqn}) 
and (\ref{Dyson.eqn}) for various values of Hubbard $U$ 
captures the physics of Mott metal-to-insulator 
transtion\cite{xyzhang93}.  
The essential many-body quantity 
provided by solving the local impurity problem is the  
self-energy which encodes the matrix element effects, as will
be shown in the following.
We solve\cite{ThanksRozenberg} the above set of equations at 
zero temperature for a Bethe lattice DOS of type
$D(\varepsilon)=\frac{2}{\pi}\sqrt{1-\varepsilon^2}$,
which corresponds to renormalized hopping $\tilde t=t\sqrt d=1/2$. 
 
   The third order nonlinear optical response per unit volume  is related to  
four-current correlation  
$\chi^{(3)}_{jj}(\omega;\omega_1,\omega_2,\omega_3)$ 
by\cite{Wu} 
\begin{small}
\begin{align} 
   \chi^{(3)}(\omega;\omega_1,\omega_2,\omega_3) &= 
   \frac{1}{3!}\left(\frac{-i}{\hbar}\right)^3 \frac{1}{V} 
   \frac{\chi^{(3)}_{jj}(\omega;\omega_1,\omega_2,\omega_3)}  
   {i \omega_\sigma\omega_1\omega_2\omega_3}, 
\end{align} 
\end{small}
where $\omega=-\omega_\sigma=-(\omega_1+\omega_2+\omega_3)$, and the  
$4-$current correlation function is given by 
\begin{align} 
\begin{split} 
   \chi^{(3)}_{jj}(\omega;\omega_1,\omega_2,\omega_3) &=\! \! 
   \int dx dx_1 dx_2 dx_3
    e^{i\omega t+\omega_1 t_1+\omega_2 t_2+\omega_3 t_3}  \nn\\
   & \times\langle T_c \jmath(x)\jmath(x_1)\jmath(x_2) \jmath(x_3)\rangle.\nn 
\end{split} 
\end{align} 
where $\jmath(x)$ is current operator at space-time $x=(\vec r,t)$. 
Here $T_c$ denotes the time-ordering along the Keldysh  
path\cite{Yu.Lu}. Although the general formulation can be 
written down in terms of Keldysh Green's functions,  
but for parametric processes\cite{BoydBook}, i.e. processes  
in which, final and initial states are identical, in practice 
one can avoid use of Keldysh Green's functions. In such a case 
one can use ordinary Green's function, to calculate the  
{\em time ordered} diagrams, followed by appropriate 
analytic continuation to ensure the correct $\nu+i\eta$ behavior 
of the fully retarded optical responses\cite{JafariJPSJ}. 
 
\begin{figure}[t] 
   \centering 
   \includegraphics[height=0.25\textwidth,angle=0]{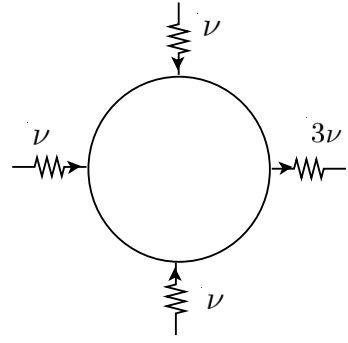} 
   \caption {Feynman diagram corresponding to the third harmonic generation.  
   This diagram represents the {\em time-ordered}  
   four-current correlation. To obtain the {\em fully retarded} four-current 
   correlation we ensure the correct $\nu+i\eta$ analytic behavior.} 
   \label{FeynmanTHG.fig} 
\end{figure}

   The case of third harmonic generation corresponds to 
$\omega_1=\omega_2=\omega_3=\nu$, so that we have 
\begin{small}
\begin{align} 
   \label{thg8.eqn}
   &\chi^{\rm THG}(\nu) \equiv\chi^{(3)}(-3\nu;\nu,\nu,\nu) =  
   \frac{1}{18}\left(\frac{1}{\hbar}\right)^3  
   \frac{\langle\jmath\jmath\jmath\jmath\rangle^{\rm THG}(\nu)}{\nu^4},\\ 
   &\langle\jmath\jmath\jmath\jmath\rangle^{\rm THG}(\nu) 
   \!\!=\!\! \int\!\! dx e^{-3i\nu t}\Pi_i dx_i 
   e^{i\nu t_i} 
   \langle T \jmath(t)\jmath(x_1)\jmath(x_2) \jmath(x_3)\rangle. \nn
\end{align} 
\end{small}
The Feynman diagram corresponding to the THG process\cite{J.Yu} is shown in 
Fig. \ref{FeynmanTHG.fig}.  
In the limit of infinite dimensions  
vertex corrections to odd parity operators indentically vanish 
by Ward indentity.  
  To see this, let us write down the Ward identity as\cite{PeskinBook} 
\begin{align} 
   -ik_\mu \Gamma^\mu(p+k,k)=G^{-1}(p+k)-G^{-1}(k) 
\end{align} 
where summation over $\mu=0,1\ldots,d$ is understood and $p,k$ are 
($d+1$)-vectors. 
Using the Dyson equation the right hand side becomes
$ \Sigma(k+p)-\Sigma(k)$. In $d\to\infty$ limit 
self-energy is purely local\cite{KotliarRMP} (no $k$ dependence) and 
hence it vanishes. Now, since the current ($\propto$velocity) 
vertex has odd parity under $\vec k\to -\vec k$,  
the vertex correction $\Gamma^\mu$ identically vanishes. 
 
Therefore the four-current correlation 
$\langle\jmath\jmath\jmath\jmath\rangle^{\rm THG}(\nu)$ 
in Fig. \ref{FeynmanTHG.fig} can be obtained by simple 
convolution. The Green's functions running around the loop 
are self-consistent lattice Green's function obtained from 
the impurity solver by iterated perturbation theory\cite{KotliarRMP}. 
Now let us further simplify 
$\langle\jmath\jmath\jmath\jmath\rangle^{\rm THG}(\nu)$ 
in $d\to\infty$ limit.
Equation (\ref{thg8.eqn}) can be written as
\begin{align} 
   &\chi^{\rm THG}(i\nu) =\frac{1}{18} 
   \frac{\langle\jmath_1\jmath_1\jmath_1\jmath_1\rangle}{\nu^4}= 
   \frac{8\tilde t^4 }{9\nu^4}\frac{1}{Nd\beta}
   \sum_{\vec k,\omega_n} \sin^4(k_1)\times \nn\\&
   G_{\vec k\sigma}(i\omega_n)G_{\vec k\sigma}(i\omega_n+i\nu) 
   G_{\vec k\sigma}(i\omega_n+2i\nu) G_{\vec 
   k\sigma}(i\omega_n+3i\nu)\nn 
\end{align}  
where we have used the fact that current vertex in 
direction no. $1$ is $2\tilde t \sin(k_1)$. To proceed further,
we need to define $\rho_0(\varepsilon)=\frac{1}{Nd}\sum_{\vec k}\sin^4(k_1)
\delta(\varepsilon-\varepsilon_{\vec k})$. To take the limit $d\to\infty$, we 
Fourier transform $\rho_0(\varepsilon)$ as:
\begin{align} 
\begin{split} 
   &\rho_0(s) = \int \rho_0(\varepsilon) e^{is\varepsilon} d\varepsilon = 
   \frac{1}{Nd}\sum_{\vec k} 
     \sin^4(k_1)e^{is\varepsilon_{\vec k}}\nn\\ 
     &=\left[\int_{-\pi}^\pi e^{-2ist\cos k}\frac{dk}{2\pi}\right]^{d-1}
     \left[\int_{-\pi}^\pi \sin^4 k_1~ e^{-2ist\cos k_1}\frac{dk_1}{2\pi}\right]\nn\\
     &=\left[J_0(2st)\right]^{d-1} \times
     \frac{1}{8}\left[3J_0(2st)-4J_2(2st)+J_4(2st)\right] \nn\\
   &= \frac{1}{8}\left[ 3 [J_0(2st)]^d+{\cal O}\left(\frac{1}{d}\right) \right], 
\end{split} 
\end{align} 
where $J_n$ is the Bessel function of order $n$.
In the last line we have used the fact that in $d\to\infty$ 
the hopping matrix element scales like $t=\tilde t/\sqrt d$ and 
hence $st\ll 1$, so that using
$J_n(x)\approx \frac{x^n}{2^nn!}$, we can ignore $J_2$ and 
$J_4$ compared to $J_0$. Repeating the above algebra without
$\sin^4 k_1$ shows that, $\left[J_0(2st)\right]^d$  
is indeed the Fourier transform of $D(\varepsilon)$. Therefore 
\begin{align} 
   \rho_0(\varepsilon) = \frac{3}{8} D(\varepsilon),
\end{align} 
which allows us to write 
\begin{align} 
   \label{thgConvolutionT.eqn} 
   &\chi^{\rm THG}(i\nu) = \frac{\tilde t^4}{3\nu^4\beta} \sum_{\omega_n}\int 
   d\varepsilon D(\varepsilon)\times \\&
   G(\varepsilon,i\omega_n) 
   G(\varepsilon,i\omega_n+i\nu)
   G(\varepsilon,i\omega_n+2i\nu) 
   G(\varepsilon,i\omega_n+3i\nu) \nn
\end{align} 
We see that in $d\to\infty$ the $\vec k$ summation becomes simply a 
DOS integration. In the following 
$\varepsilon$ stands for $\varepsilon_{\vec k}$, and the explicit $\varepsilon$ 
subscript emphasises the $\vec k$ label. From the derivation it 
can be seen that other four current correlations like 
$\langle\jmath_1\jmath_1\jmath_2\jmath_2\rangle$ in $d\to\infty$ limit
differ from $\langle\jmath_1\jmath_1\jmath_1\jmath_1\rangle$ 
by a numerical factor. Therefore to that extent the limit of $d\to\infty$
is blind to various directions in space. Hence DMFT
can not distinguish optical spectroscopies with polarized light 
from those with unpolarized light.
 
  To elucidate the matrix element effects in DMFT method, 
after Lehman representing the Greens' functions in terms of
$A(\vec k,E)\to A(\varepsilon,E)$, and using  standard contour integration 
techniques to perform $1/\beta\sum_{\omega_n}$ summation we obtain
\begin{align} 
\begin{split} 
   &\chi^{\rm THG}(i\nu) = -\frac{\tilde t^4}{3\nu^4} 
   \int d\varepsilon D(\varepsilon) dE_1\ldots dE_4\\ 
   &A(\varepsilon,E_1) A(\varepsilon,E_2) 
   A(\varepsilon,E_3) A(\varepsilon,E_4)\times F,
\end{split} 
\end{align} 
with
\begin{align} 
\begin{split} 
   F&=\frac{f(E_1)}{E_1-E_2+i\nu}\frac{1}{E_1-E_3+2i\nu}\frac{1}{E_1-E_4+3i\nu}\\ 
   &+\frac{f(E_2)}{E_2-E_1-i\nu}\frac{1}{E_2-E_3+i\nu}\frac{1}{E_2-E_4+2i\nu}\\ 
   &+\frac{f(E_3)}{E_3-E_1-2i\nu}\frac{1}{E_3-E_2-i\nu}\frac{1}{E_3-E_4+i\nu}\\ 
   &+\frac{f(E_4)}{E_4-E_1-3i\nu}\frac{1}{E_4-E_2-2i\nu}\frac{1}{E_4-E_3-i\nu} 
   \label{thgLehman.eqn} 
\end{split} 
\end{align} 
where $f$ is the Fermi function.
This expression closely resembles familiar expressions 
in nonlinear optics literature (see e.g. Sec. 3.2  
of Ref. 2). Therefore in this formulation, 
the matrix element effects  
appear via spectral function $A(\varepsilon,E)$, which 
itself is fully determined by the self-energy. In principle
after replacing $i\nu$ with $\nu+i 0^+$ in this expression, we
can use the spectral weights obtained from DMFT solver to calculate
the nonlinear response. However, numerical calculation of the above
five dimensional integrals is not computationally feasible.

   Another alternative method would be to calculate $\chi^{\rm THG}$ at
Matsubara frequencies according to
(\ref{thgConvolutionT.eqn}), followed by analytic
continuation $i\nu\to \nu+i0^+$. But in this process, we  face
spurious features characteristic of analytic continuation of numbers,
which makes it difficult to assess the nonlinear dynamical 
structures. 

   Since we are interested in high energy features in the scale of Mott gap, 
which is much larger than the thermal energies at room temperature,  
in order to avoid the above mentioned difficulties, 
we work at zero temperature. At $T=0$, 
(\ref{thgConvolutionT.eqn}) will be replaced by 
\begin{align} 
   \chi^{\rm THG}(\nu) \!=\!  
   \frac{\tilde t^4}{6\pi\nu^4} \!\int\!  
   \frac{d\omega d\varepsilon D(\varepsilon)}{\xi_0-\varepsilon} 
   \frac{1}{\xi_1-\varepsilon} 
   \frac{1}{\xi_2-\varepsilon} 
   \frac{1}{\xi_3-\varepsilon}, 
   \label{thgConvolution.eqn}
\end{align}
where 
$\xi_j=\omega+j\nu-\Sigma_R(\omega+j\nu)+i\vert\Sigma_I(\omega+j\nu)\vert$ 
for $j=0,1,2,3$.
 
   In the above formula, (i) the integration over $\varepsilon$  
corresponds to summation over intermediate states in conventional 
expressions\cite{BoydBook} which are usually used for 
systems with discrete energy levels, and 
(ii) the matrix element effects are encoded in $\Sigma(\omega)$, 
real and imaginary part of which have been denoted by $\Sigma_R$ 
and $\Sigma_I$, respectively. 
{\em It is very crucial to note that we have used absolute value  
of the imaginary part of the self-energy}. This is indeed 
a necessary step to transform from time-ordered four-current, 
to fully retarded one\cite{JafariJPSJ}. 

   Decomposing the integrand in (\ref{thgConvolution.eqn}) to 
partial fractions in terms of $\varepsilon$, the four-current correlation
can be written as $f_0+f_1+f_2+f_3$, where 
\begin{align}
   f_j(\nu)=\frac{\tilde t^4}{6\pi}\int d\omega
   \prod_{s\ne j}\frac{1}{(\xi_j-\xi_s)}
   \int d\varepsilon\frac{D(\varepsilon)}{(\varepsilon-\xi_j)}
   \label{decompose.eqn}
\end{align}
The expression (\ref{decompose.eqn})
reveals the resonance structure transparently: It arises from 
$\omega+j\nu-\Sigma_R(\omega+j\nu)=\varepsilon$. When
the frequency $\nu$ of the photons is such that 
$j$-photon frequency matches the energy difference 
$\varepsilon-\omega+\Sigma_R(\omega+j\nu)$, we will have
a $j$-photon resonance. 
Here, $f_0$ corresponds to a background contribution. 

  Note that within this formulation, 
we do not have any control over the 'broadening'
parameter $\eta$, instead the 
broadening $\eta=\vert\Sigma_I(\omega)\vert$ is determined by the
solution of the impurity problem in a self-consistent fashion.
\begin{figure}[t] 
   \centering 
   \includegraphics[height=0.35\textwidth,angle=0]{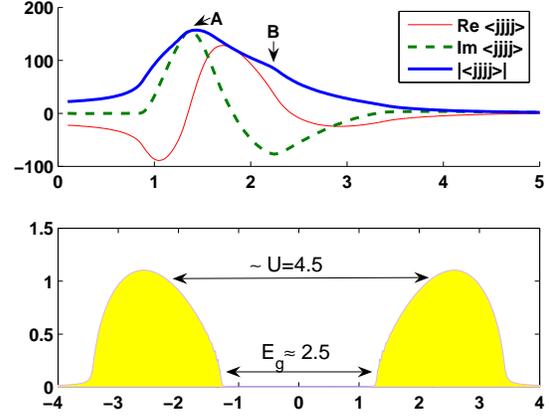} 
   \caption {(Color online) THG and DOS for $U=4.5$. We use the 
   Bethe lattice in solving DMFT equations. In the upper pannel, dashed
   line is the imaginary part of $\langle \jmath\jmath\jmath\jmath\rangle$, 
   narrow solid line represents its real part, and the bold solid line represents
   the absolute value. Lower pannel shows the DOS. The onset of (3-photon)
   absorption at $\nu\approx 0.85$ corresponds to the gap value,
   $E_g\approx2.5$, while peak-to-peak energy difference is
   on the scale of $U=4.5$}
   \label{thgU4p5.fig} 
\end{figure} 

Now let us present our results.
For semicircular Bethe lattice DOS of width $2\tilde t=1$, the
critical value is given by $U_c\approx 3.3$, above which system is in the
Mott insulating state. Fig. \ref{thgU4p5.fig} shows the result
for $U=4.5$. Lower pannel shows the self-consistent DOS, with a 
Mott-Hubbard gap $\sim 2.5$. The peak-to-peak separation between
the upper and lower Hubbard bands is $\sim U=4.5$. Upper pannel shows
real(dashed), imaginary(dotted) and absolute value(solid line) of the 
four-current correlation 
$\langle \jmath\jmath\jmath\jmath\rangle^{\rm THG}$.

   The onset of absorption starts at $\nu\approx0.85$ which is
$1/3$ of the gap magnitude. This can be clearly seen in 
the imaginary part of the THG four-current correlation
in Fig. \ref{thgU4p5.fig}. This onset clearly corresponds
to three-photon absorption. The three-photon resonance 
peaks around $\nu\approx 1.5$ (denoted by {\bf A}) which 
is $1/3$ of the peak-to-peak separation of the Hubbard bands.
Next weaker feature (denoted by {\bf B}), which is a shoulder
similar to finite-dimensional results\cite{KishidaPRL},
corresponds to $1/2$ of peak-to-peak 
separation ($\sim 4.5$) of the Hubbard bands. The one-photon process is the
weakest feature around $\nu \sim 4.5$ which can hardly be distinguished in THG
spectrum. Despite that DMFT is designed to work better in larger dimensions, 
the spectrum in figure \ref{thgU4p5.fig} qualitatively resembles the 
experimental result on Sr$_2$CuO$_3$, which is a one-dimensional
Mott insulator\cite{KishidaPRL}.

\begin{figure}[t] 
   \centering 
   \includegraphics[height=0.37\textwidth,angle=0]{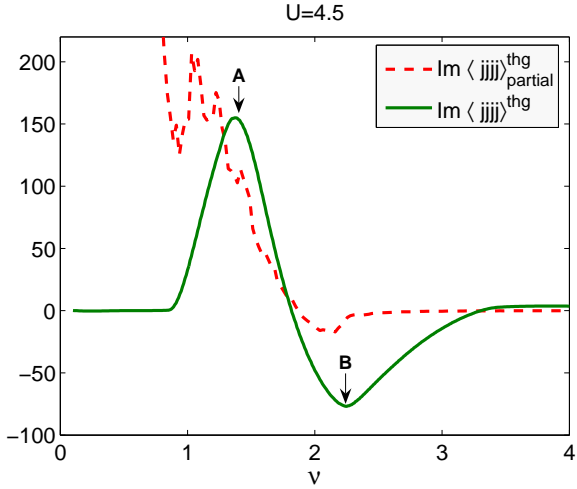} 
   \caption {Solid line shows the imaginary part of the total THG for 
   $U=4.5$. The dashed line shows the imaginary part of $f_3(\nu)$. 
   {\bf A} and {\bf B} in this figure correspond to those in 
   Fig. \ref{thgU4p5.fig}.}
   \label{thg-partial-U4p5.fig} 
\end{figure} 

   We claimed that in peak {\bf A} of Fig. \ref{thgU4p5.fig} 
the dominant contribution comes from 
three-photon processes. To demonstrate this, in Fig. 
\ref{thg-partial-U4p5.fig} we plot with dashed line, the imaginary
part of $f_3(\nu)$. The 
solid line shows the total four-current correlation. 
As can be seen, the $3-$photon resonance
dominantly contributes to the peak {\bf A}, although the peak
position is slightly shifted to lower energies. Further,
it is clearly seen that $f_3(\nu)$ deos not contribute much to the 
two photon resonance at {\bf B}. $f_3(\nu)$ also
blows up at small frequencies which by (\ref{decompose.eqn})
will be finally compensated by other patial spectra,
$f0,f1,f2$, to give the total THG spectrum. 

\begin{figure}[bht] 
   \centering 
   \includegraphics[height=0.33\textwidth,angle=0]{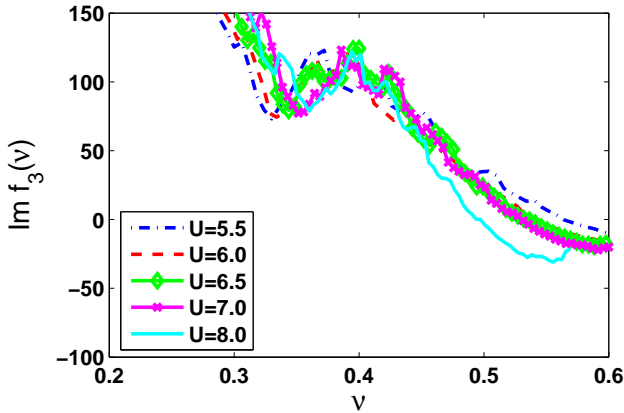} 
   \caption {Imaginary part of $f_3(\nu)$ vs. 
   $\nu/E_g$, where $E_g$ is the Mott-Hubbard gap in the Mott insulating phase.}
   \label{scaling.fig} 
\end{figure} 

In our previous work\cite{JafariJPSJ}, we found that within the
SDW mean field the THG in e.g. 1D is given by 
 $\Im \chi^{\rm THG} (\nu)=\frac{27}{2}f(\frac{3\nu}{2})
   -8f(\nu)+\frac{1}{2}f(\frac{\nu}{2})$, in which
\begin{align}
    f(\nu)&=\Delta\times\left[
    \frac{\pi}{24|\tilde \nu|^5}\left(\tilde w^2-\tilde \nu^2\right)^{3/2}
    \left(\tilde\nu^2-1\right)^{1/2}\right]
\end{align}
where the gap $E_g=2\Delta$, $\tilde w^2=(1+\Delta^2)/\Delta^2$,
and {\em the frequency $\tilde\nu=\frac{\nu}{\Delta}$ is 
scaled by the gap parameter.} In the above equation $f(3\nu/2)$ corresponds
to $f_3(\nu)$ defined in (\ref{decompose.eqn}).
We see that in the mean field approximation, $f_3(\nu)$ has a scaling part
which is universal, and independent of the gap magnitude.

Motivated by this observation, in Fig. \ref{scaling.fig}
we plot Im$f_3$ for Mott insulators with various
values of $U$ as a function of $\frac{\nu}{E_g}$. 
We also apply an overal scaling to the curves.
Such a scaling behavior, though approximate, points to a universal
features in the nonlinear optical spectra of the Mott insulators, which are
independent of the gap magnitude. It seems that the
mean field scaling behavior survives the quantum fluctuations implemented
via DMFT. It would be interesting to further explore
this observation using alternative methods of dealing with correlated
insulators.

In summary, on the technical side, 
we have formulated the nonlinear optical response theory
within the DMFT theory.
In equation (\ref{thgConvolution.eqn}) we present
a feasible way that avoids numerical difficulties.
On the physics side, assuming that DMFT is a good 
approximation for $d>1$, we observe that nonlinear
optical spectra of higher dimensional Mott insulators
share common features with  those observed in $d=1$ dimensional 
Mott insulator, Sr$_2$CuO$_3$\cite{KishidaPRL}.
Also within DMFT we observe an approximate scaling behavior
in the THG spectra of Mott insulators.

S.A.J. was supported by JSPS fellowship P04310.
We wish to thank M.J. Rozenberg for using his code in solving
DMFT equations. This work was supported by Grant-in-Aid for 
Scientific Research from the MEXT, and NAREGI project.

\end{document}